%% file: eprint.tex
\documentclass[12pt]{article}
\usepackage{graphicx}
\usepackage{subfigure}

\newcommand\pubnumber{SNSN-323-63}
\newcommand\pubdate{\today}
\def\institute{Physikalisches Institut\\
	Rheinische Friedrich-Wilhelms-Universit{\"a}t Bonn, 53115 Bonn, Germany}
\def\support{\footnote{E-mail: pienpen.seema@cern.ch}}

\def\Title#1{\begin{center} {\Large #1 } \end{center}}
\def\Author#1{\begin{center}{ \sc #1} \end{center}}
\def\Address#1{\begin{center}{ \it #1} \end{center}}

\newcommand\pubblock{\rightline{\begin{tabular}{l} \pubnumber\\
         \pubdate  \end{tabular}}}
\newenvironment{Abstract}{\begin{quotation}  }{\end{quotation}}
\newenvironment{Presented}{\begin{quotation} \begin{center} 
             PRESENTED AT\end{center}\bigskip 
      \begin{center}\begin{large}}{\end{large}\end{center} \end{quotation}}

\input econfmacros.tex

\begin{document}
\begin{titlepage}
\pubblock

\vfill
\Title{Measurement of the $t$­-channel single top-­quark and top­-antiquark differential cross-sections in $pp$ collisions at $\sqrt{s} = 8~\textrm{TeV}$ with the ATLAS detector}
\vfill
\Author{ Pienpen Seema\support \\
	on behalf of the ATLAS collaboration}
\Address{\institute}
\vfill
\begin{Abstract}
 In this document, differential cross-section measurements of $t$-channel single top­-quark production are presented. 20.2~fb$^{-1}$ of data collected by the ATLAS experiment in proton-proton collisions in the LHC at a centre­-of-­mass energy of 8~TeV are used. Differential cross-sections as a function of the transverse momentum and rapidity of both the top quark and the top antiquark have been measured at both parton and particle level. The transverse momentum and rapidity differential cross-sections of the scattered light-quark jet have been measured at particle level. All measurements are compared to different Monte Carlo predictions as well as to available fixed-­order QCD calculations.
\end{Abstract}
\vfill
\begin{Presented}
$9^{th}$ International Workshop on Top Quark Physics\\
Olomouc, Czech Republic,  September 19--23, 2016
\end{Presented}
\vfill
\end{titlepage}
\def\thefootnote{\fnsymbol{footnote}}
\setcounter{footnote}{0}

\section{Introduction}
At the LHC, top quarks can be produced singly via electroweak interactions. In leading-order perturbation theory, there are three different single top-quark production mechanisms. They are distinguished according to the virtuality of the exchanged $W$ boson. The dominant process is the $t$-channel production where a light quark interacts with a bottom quark by exchanging a space-like $W$ boson. (The other two processes are $s$-channel in which a time-like $W$ boson is exchanged and $Wt$-channel where a top quark is produced in association with a $W$ boson.) At next-to-leading order (NLO), the total cross-sections of top-quark and top-antiquark production in the $t$-channel at $\sqrt{s} = 8~\textrm{TeV}$, calculated with \textsc{HatHor}~v2.1~\cite{Kant:2014oha}, are predicted to be  
\begin{eqnarray*}
	\sigma(tq)       & = & 54.9 ^{+2.3}_{-1.9}\ \mathrm{pb}, \\
	\sigma(\bar{t}q) & = & 29.7 ^{+1.7}_{-1.5}\ \mathrm{pb}, \\
	\sigma(tq+\bar{t}q) & = & 84.6 ^{+3.9}_{-3.4}\ \mathrm{pb}.
\end{eqnarray*}

\section{Measurement}
$pp$ collisions at $\sqrt{s} = 8~\textrm{TeV}$ corresponding to an integrated luminosity of 20.2~fb$^{-1}$ recorded with the ATLAS detector~\cite{PERF-2007-01} at the LHC are used in this analysis. The signature of $t$-channel candidate events contains one charged lepton (electron or muon), high missing transverse momentum, two hadronic jets with high transverse momentum. Exactly one of the jets is required to be a $b$-jet. $t$-channel events are simulated using \textsc{Powheg-Box} ~\cite{Frederix:2012dh} generator with \textsc{Pythia6}~\cite{Sjostrand:2006za} for parton shower and hadronisation.

The measurements are performed at both particle level and parton level. A top quark reconstructed from stable particles at particle level in a fiducial phase space is called a pseudo-top-quark, $\hat{t}$. Similarly, a scattered light jet at particle level is written as $\hat{j}$. At parton level, a truth top quark, $t$, is defined as a top quark after gluon radiation. The parton-level top quark is theoretically well defined, thus the measurement can be compared to theory predictions directly.  

Both absolute and normalised differential cross-sections are measured in bins of $p_{\rm{T}}(\hat{t})$, $|y(\hat{t})|$, $p_{\rm{T}}(\hat{j})$ and $|y(\hat{j})|$ at particle level as well as of $p_{\rm{T}}(t)$ and $|y(t)|$ in the full phase space at parton level. A neural network (NN) is utilised to discriminate signal from background events. Good separation is achieved by using seven input variables where $|\eta{(j)}|$ is sorted as the second most powerful variable. This NN is used for all measurements except for the measurement as a function of $|y(\hat{j})|$, where a second NN is trained without $|\eta{(j)}|$. All differential cross-sections are extracted in a signal enriched region where a cut on the NN output of 0.8 is applied.   
		
\begin{figure}[!h!tpb]
	\centering
	\subfigure[]{
		\includegraphics[width=0.45\textwidth]{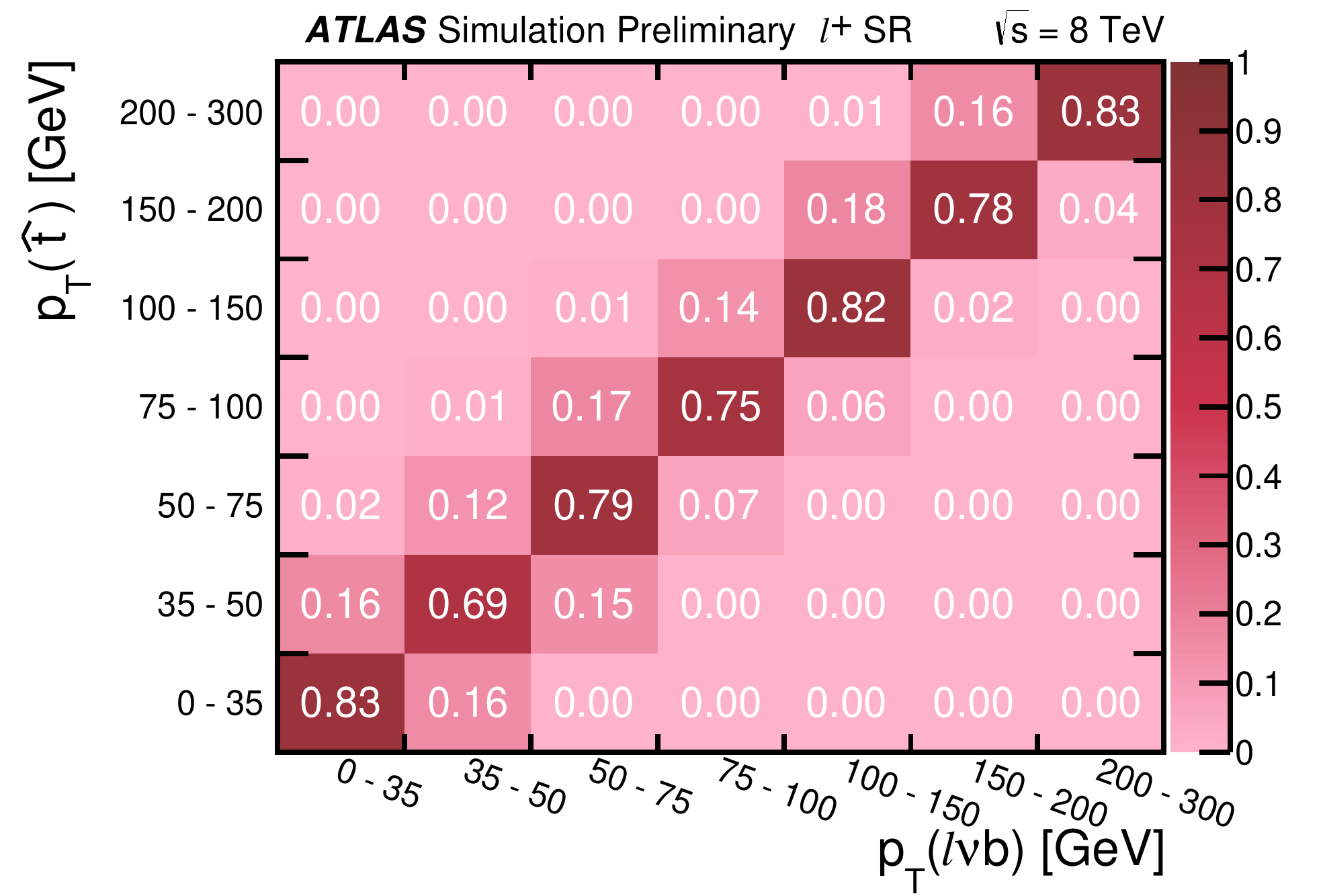}
	}
	\subfigure[]{
		\includegraphics[width=0.45\textwidth]{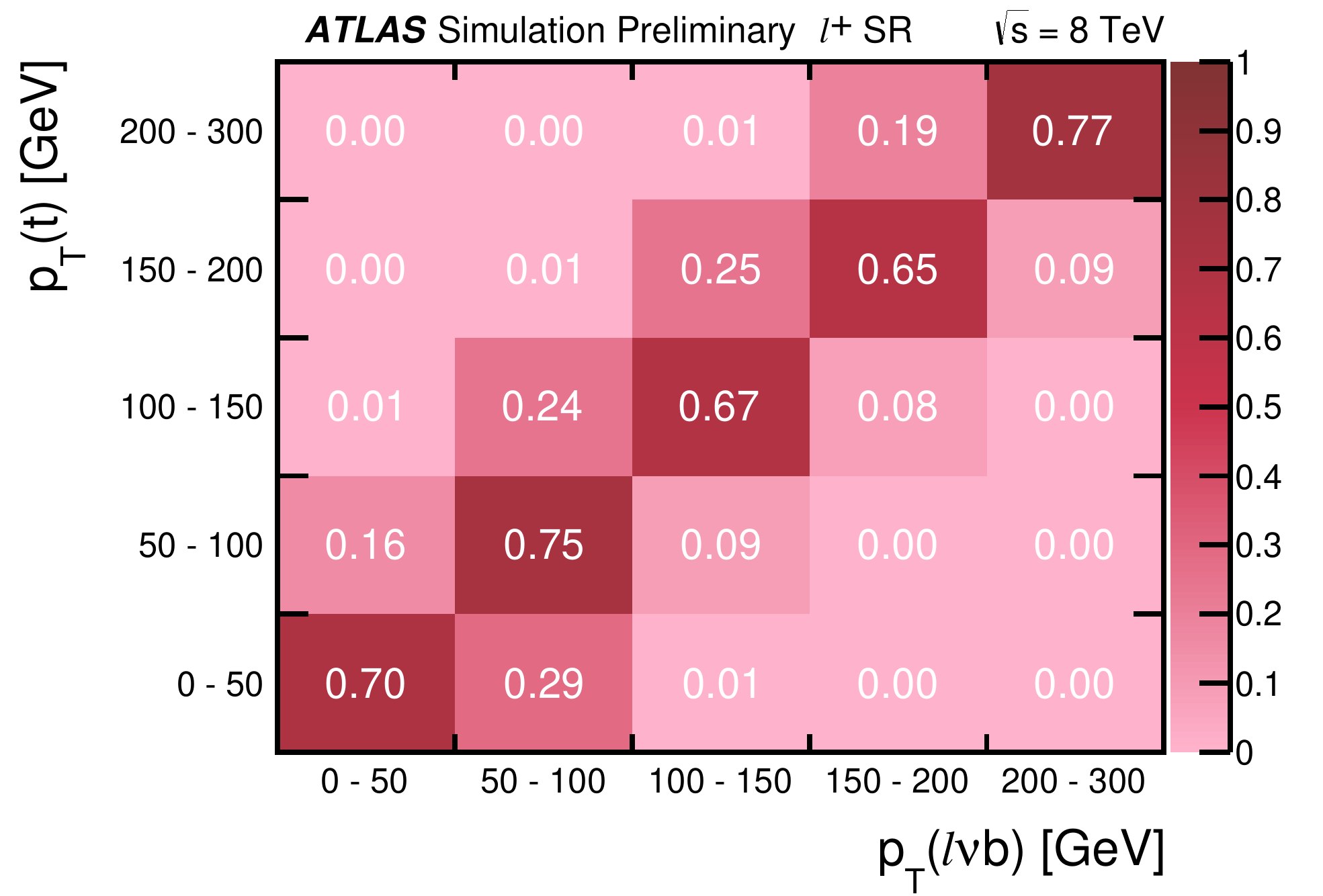}
	}
	\caption{Migration matrices for (a) $p_{\rm{T}}(\hat{t})$ at particle level and (b) $p_{\rm{T}}(t)$ at parton level~\cite{paper}.
		The particle level or parton level is shown on the $y$-axis and the reconstructed level is shown on the $x$-axis.}
	\label{fig:mm}
\end{figure}

Measured observables are distorted by detector effects and selection efficiency. An iterative Bayesian unfolding method~\cite{DAgostini1995487}, implemented inside RooUnfold framework~\cite{Adye:2011gm} is exploited in order to correct for these effects. Figure~\ref{fig:mm} shows migration matrices for $p_{\rm{T}}(\hat{t})$ at particle level and $p_{\rm{T}}(t)$ at parton level. The particle-level migration matrix has much smaller off-diagonal terms due to better resolution of the pseudo-top-quark. 

\hspace{1 cm}

\section{Results}
\begin{figure}[!h!tpb]
	\centering
	\subfigure[]{
		\includegraphics[trim={0 18pt 0 37pt}, clip, width=0.37\textwidth]{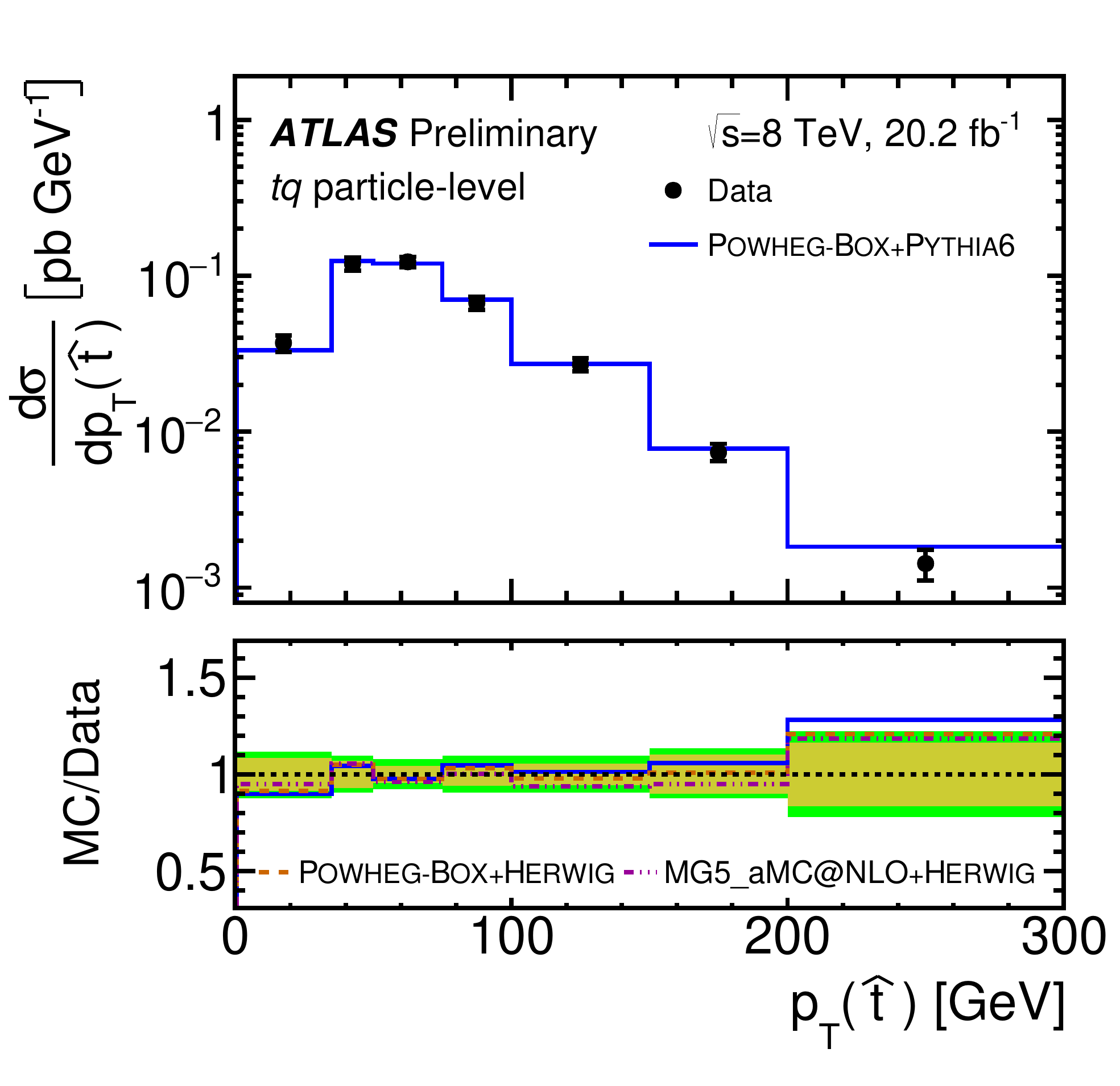}
	}
	\subfigure[]{
		\includegraphics[trim={0 18pt 0 37pt}, clip, width=0.37\textwidth]{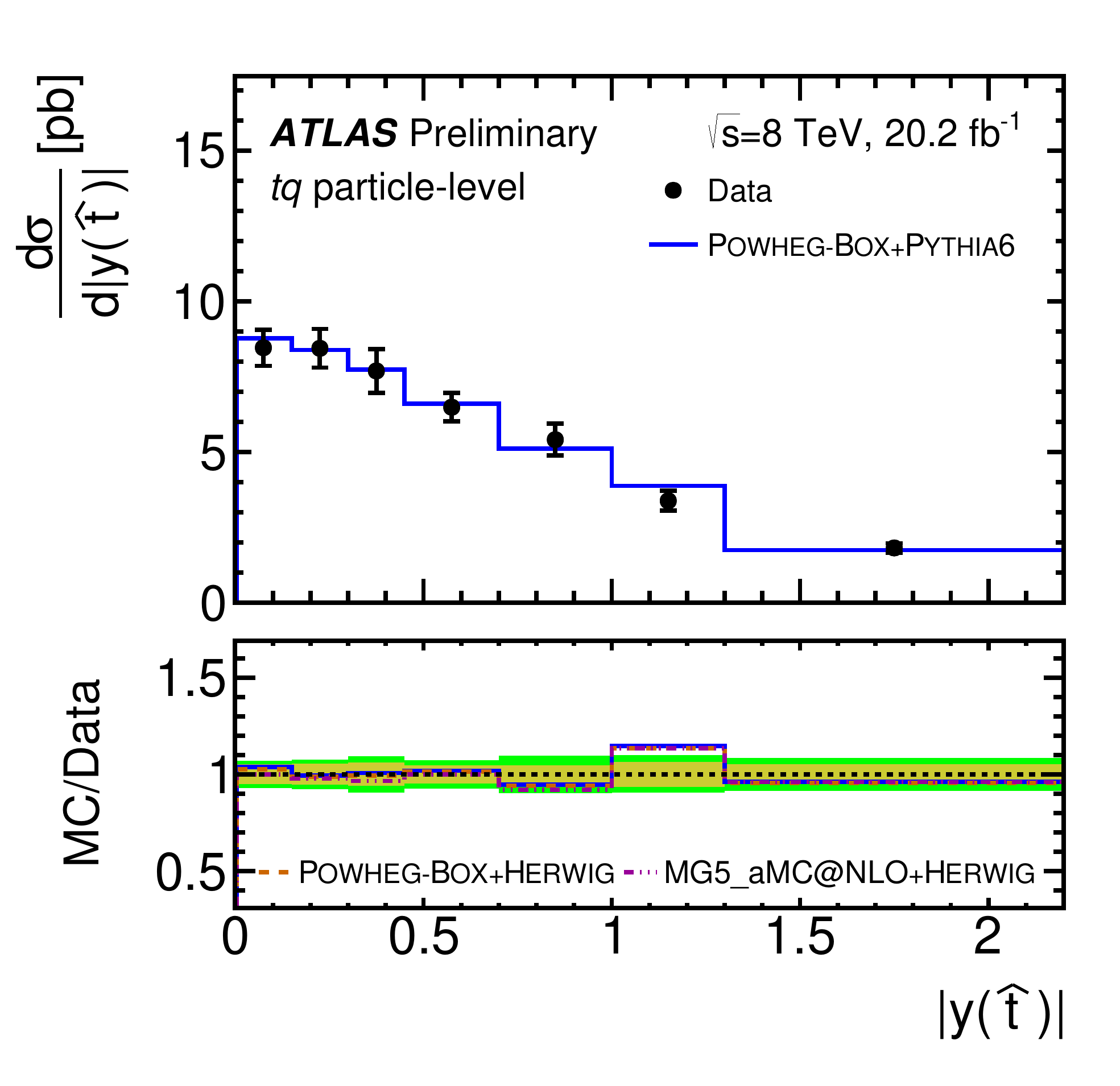}
	}
	\subfigure[]{
		\includegraphics[trim={0 18pt 0 37pt}, clip, width=0.37\textwidth]{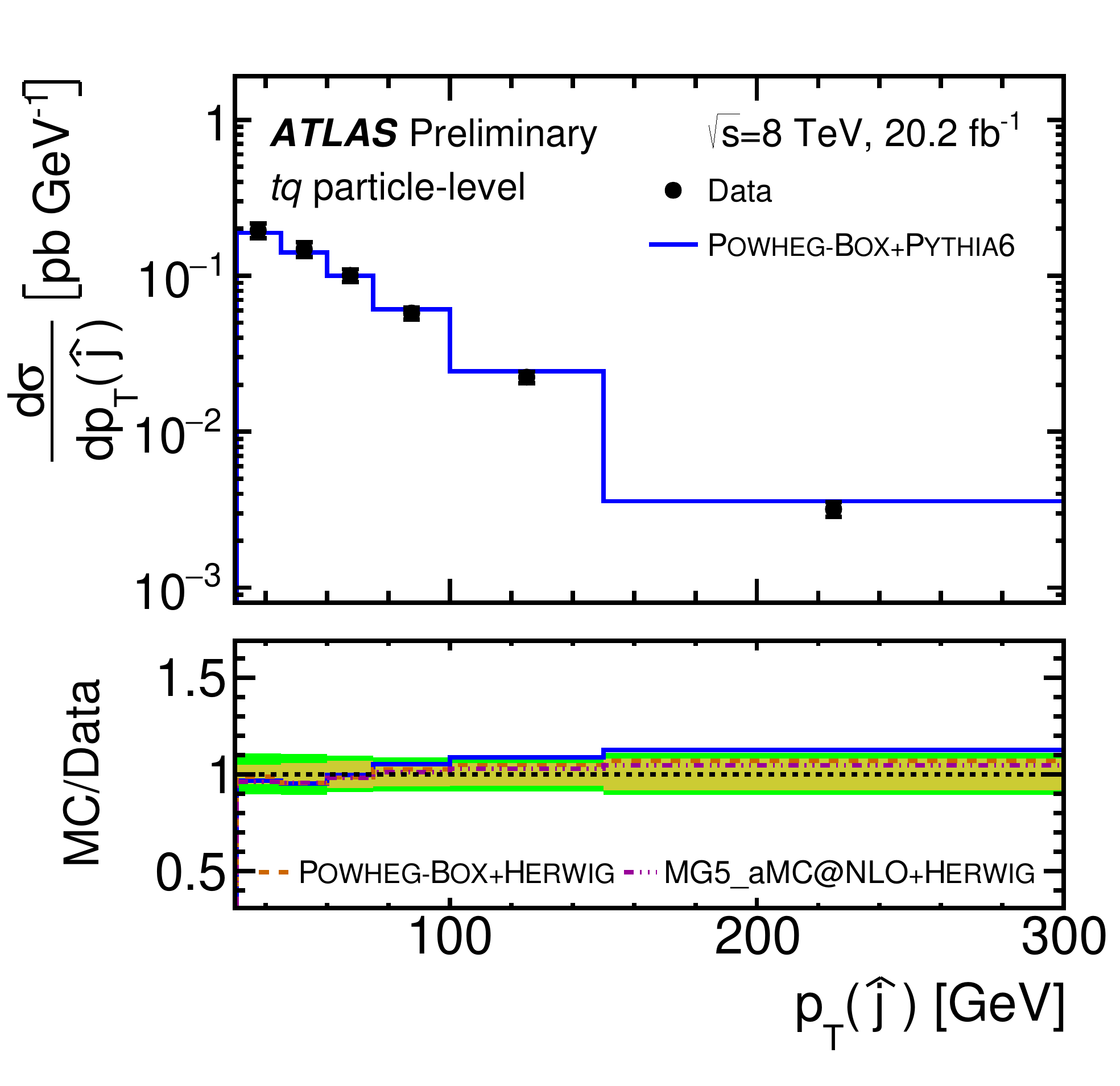}
	}
	\subfigure[]{
		\includegraphics[trim={0 18pt 0 37pt}, clip, width=0.37\textwidth]{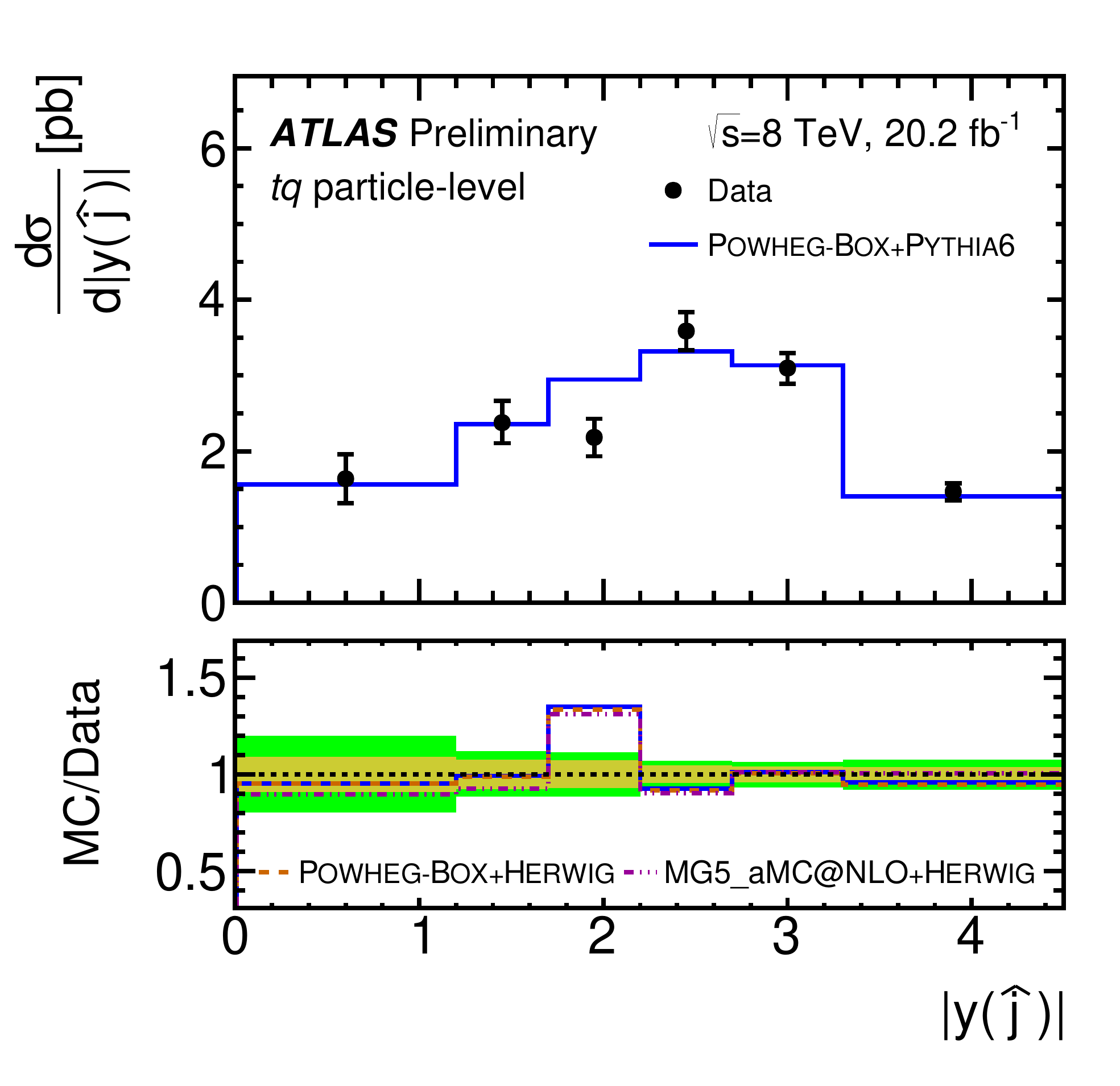}
	}	
	\subfigure[]{
		\includegraphics[trim={0 18pt 0 37pt}, clip, width=0.37\textwidth]{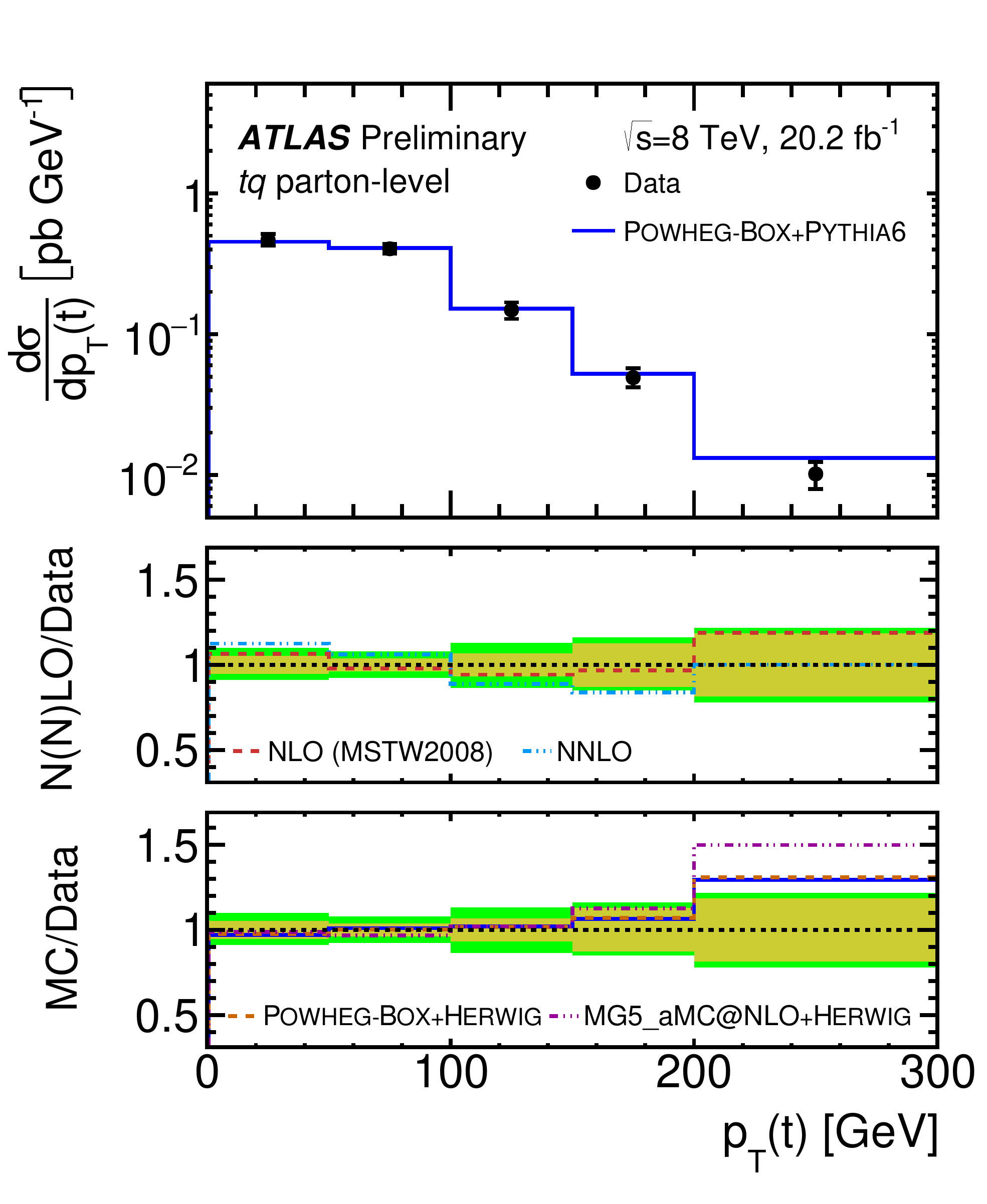}
	}
	\subfigure[]{
		\includegraphics[trim={0 18pt 0 37pt}, clip, width=0.37\textwidth]{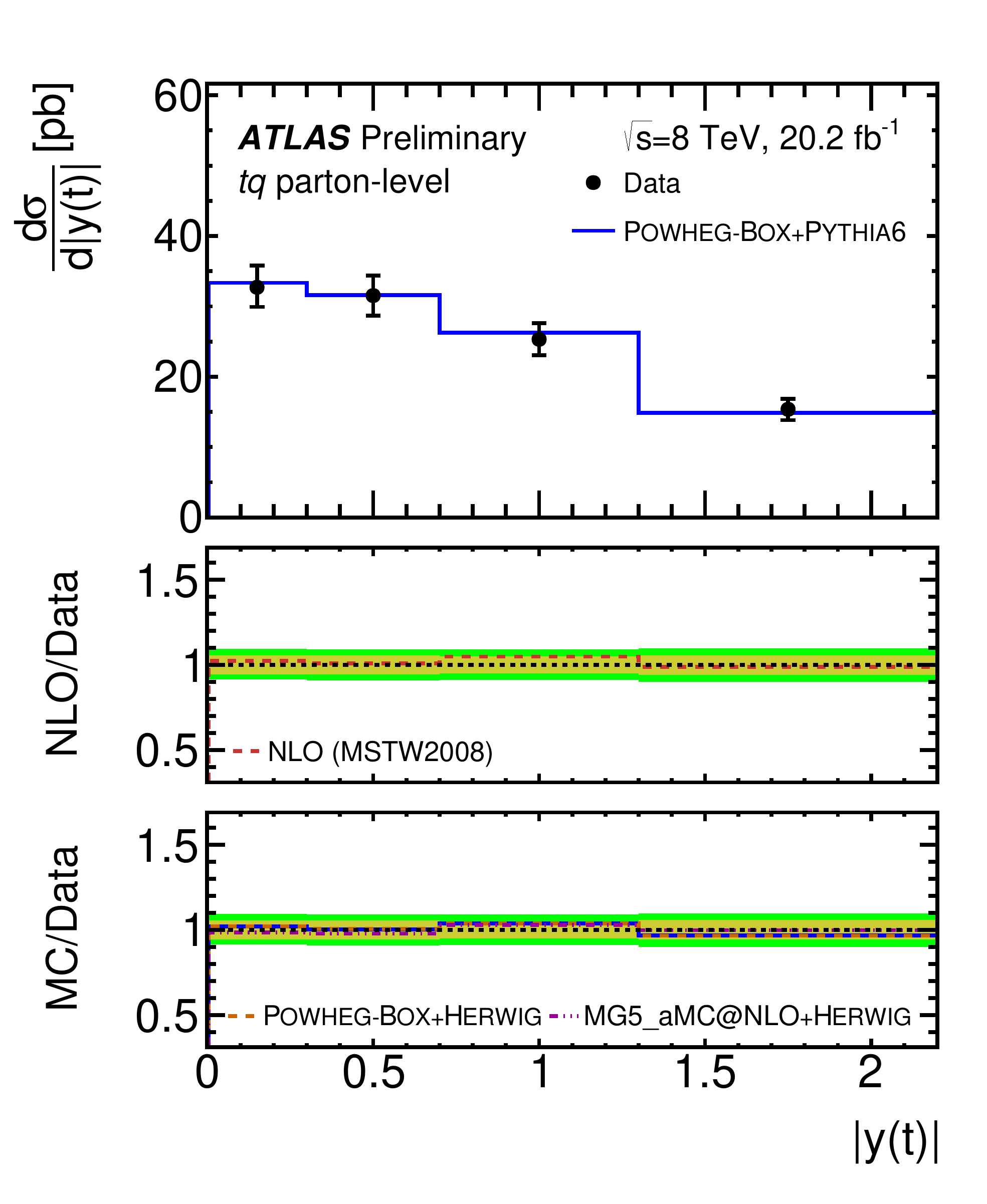}
	}	
	\caption{Absolute $t$-channel single top-quark differential cross-sections as a function of (a) $p_{\rm{T}}(\hat{t})$, (b) $|y(\hat{t})|$, (c) $p_{\rm{T}}(\hat{j})$, (d) $|y(\hat{j})|$ (e) $p_{\rm{T}}(t)$ and (f) $|y(t)|$~\cite{paper}. The distributions are compared to MC predictions (and fixed-order QCD calculations). The inner (outer) error band represents the statistical (total) uncertainty.}
	\label{fig:ptcl}
\end{figure}

Figure~\ref{fig:ptcl} shows a selection of the results at particle level and at parton level. Particle-level cross-sections are compared to MC predictions using the \textsc{Powheg-Box} and \textsc{Madgraph5}\_aMC@NLO~\cite{Alwall:2014hca} generators. Comparisons to different parton-shower and hadronisation models: \textsc{Pythia6} or \textsc{Herwig}~\cite{Corcella:2000bw} interfaced to \textsc{Powheg-Box} are also shown. 
All measured differential cross-sections agree well with the MC predictions, even though the predicted spectra tend to be slightly harder than the data for the measurement as a function of $p_{\rm{T}}(\hat{j})$.

Parton-level cross-sections are confronted with the same MC predictions and NLO predictions calculated using MCFM~\cite{Campbell}. A comparison to a calculation at approximate NNLO from Kidonakis is available for $p_{\rm{T}}(t)$~\cite{Kidonakis:2013yoa}. Good agreement is achieved between the data and all predictions. The data is described better by the approximate NNLO prediction than the MC predictions as a function of $p_{\rm{T}}(t)$.

The overall uncertainty is typically $5\%-10\%$ precision per bin. The main contributions are from the JES calibration, the signal modelling and the $t\bar{t}$\ background.
In general, the total systematic uncertainty for the normalised differential cross-sections is smaller than the uncertainty for the absolute differential cross-sections because many systematic uncertainties are reduced or cancelled for the normalised cross-section measurements.

%
%

\section{Conclusion}
Measurements of absolute and normalised $t$-channel single top-(anti)quark differential cross-sections with the ATLAS detector using 20.2~fb$^{-1}$ of $pp$ collision data at $\sqrt{s} = 8~\textrm{TeV}$ are presented. The analysis is performed in the lepton+2 jets channel. Differential cross-sections as a function of the transverse momentum and absolute value of the rapidity of the top quark, the top antiquark, and the scattered light jet are performed  at particle level for the first time. Differential cross-sections as a function of the transverse momentum and absolute value of the rapidity of the top quark and top antiquark are also performed at parton level. The most precise measurements are the normalised differential cross-sections at particle level. All measured differential cross-sections are well described by different Monte Carlo predictions as well as by available fixed-order QCD calculations.

\end{document}

%% file: econfmacros.tex
\def\beq{\begin{equation}}
\def\eeq#1{\label{#1}\end{equation}}
\def\eeqn{\end{equation}}

\def\beqa{\begin{eqnarray}}
\def\eeqa#1{\label{#1}\end{eqnarray}}
\def\eeqan{\end{eqnarray}}

\let\bar=\overbar

\def\Dslash{\not{\hbox{\kern-4pt $D$}}}
\def\dslash{\not{\hbox{\kern-2pt $\del$}}}

\def\msb{{\bar{\ssstyle M \kern -1pt S}}}

